\documentclass[reprint,aps,prd,superscriptaddress,showkeys,showpacs]{revtex4-1}
\usepackage{epsfig,amsmath,natbib}

\usepackage{aas_macros}
\usepackage{amssymb}
\usepackage{amsmath}
\usepackage{dsfont}
\usepackage{hyperref}
\usepackage{color}
\usepackage{pbox}
\usepackage{enumerate}

\hypersetup{
	colorlinks=false,
	citecolor=green
}

\begin{document}

\title{Geometry and growth contributions to cosmic shear observables}

\author{Jos\'e Manuel Zorrilla Matilla}
\email{jzorrilla@astro.columbia.edu}
\affiliation{Department of Astronomy, Columbia University, New York, NY 10027, USA}
\author{Zolt\'an Haiman}
\affiliation{Department of Astronomy, Columbia University, New York, NY 10027, USA}
\author{Andrea Petri}
\affiliation{Department of Physics, Columbia University, New York, NY 10027, USA}
\author{Toshiya Namikawa}
\affiliation{Department of Physics, Stanford University, Stanford, California 94305, USA}
\affiliation{Kavli Institute for Particle Astrophysics and Cosmology, SLAC National Accelerator Laboratory, 2575 Sand Hill Road, Menlo Park, California 94025, USA}

\date{\today}

\begin{abstract}
We explore the sensitivity of weak lensing observables to the
expansion history of the Universe and to the growth of cosmic
structures, as well as the relative contribution of both effects to
constraining cosmological parameters. We utilize ray-tracing
dark-matter-only N-body simulations and validate our technique by
comparing our results for the convergence power spectrum with analytic
results from past studies. We then extend our analysis to non-Gaussian
observables which cannot be easily treated analytically.
We study the convergence (equilateral)
bispectrum and two topological observables, lensing peaks and
Minkowski functionals, focusing on their sensitivity to the matter
density $\Omega_m$ and the dark energy equation of state $w$. We find
that a cancelation between the geometry and growth effects is a common
feature for all observables, and exists at the map level. It
weakens the overall sensitivity by up to a factor of 3 and 1.5
for $w$ and $\Omega_m$, respectively, with the bispectrum worst
affected. However, combining geometry and growth information alleviates the
degeneracy between $\Omega_m$ and $w$ from either effect alone.  As a
result, the magnitude of marginalized errors remain similar to those
obtained from growth-only effects, but with the correlation between
the two parameters switching sign.  These results shed light on the
origin of cosmology-sensitivity of non-Gaussian statistics, and should
be useful in optimizing combinations of observables.
\end{abstract}

\maketitle

\section{introduction}

A cosmological model with a nearly scale-invariant primordial
fluctuation spectrum, cold dark matter (CDM) and dark energy (DE)
matches well a wide range of observations, from the Universe's
expansion measured by standard candles \cite{Riess98, Perlmutter99}
and standard rulers \cite{Anderson14}, to its primordial chemical
composition \cite{ABG, Copi95}, structure formation and the properties
of the Cosmic Microwave Background (CMB) \cite{Planck15}. While
non-baryonic DM and DE make up most of the present-day energy density
of the Universe \cite{Fukugita04}, the nature of either dark component
remains unclear.

Cosmic shear is the weak gravitational lensing of background sources
by large scale structure \cite{Bartelmann01, Kilbinger15}. It probes the
matter density field through the gravitational potential fluctuations,
and is also sensitive to the expansion history of the Universe through
the distances between the observer, lensed source and lensing
structures. While lensing is usually characterized by a measurement of
the shear through the shapes of background galaxies, convergence (magnification) 
statistics can be inferred from these measurements, and are considered here for ease of 
computation. The polyspectra of the convergence field are equal to the E-modes of the shear field.

Ongoing and upcoming surveys, such as the Dark Energy Survey
(DES \footnote{\url{http://www.darkenergysurvey.org}}), the Large
Synoptic Survey Telescope (LSST \footnote{\url{http://www.lsst.org}}),
the Euclid mission \footnote{\url{http://sci.esa.int/Euclid/}} and the
Wide Field Infrared Survey Telescope
(WFIRST \footnote{\url{wfirst.gsfc.nasa.gov}}), include weak lensing
in their scientific program as part of their effort to test the
concordance model with unprecedented precision and shed light on the
nature of DM and DE. To realize this potential, we need observables
that extract all the cosmological information from the data, as well as
models capable of predicting them with high accuracy.

Second-order statistics do not fully capture non-Gaussianities
in the lensing signal from non-linear gravitational collapse
on small scales. Numerous alternative observables have been proposed
to extract this extra information, from higher-order correlation
functions \cite{Bernardeau97, Hui99} and moments \cite{Munshi01} to
topological features like local maxima (peaks) \cite{Jain00} and
Minkowski functionals \cite{Sato01}.

In this work, our goal is to clarify the sensitivity of such observables
to the expansion history of the Universe (``geometry'') and to the
evolution of primordial inhomogeneities into cosmic structures
(``growth''). The analogous question has been addressed for
the convergence ($\kappa$) power spectrum \cite{Simpson05}. The
geometry {\it vs.} growth decomposition of the power spectrum
has improved our understanding of constraints on
DE from weak lensing \cite{Zhan09}, provided an alternative
cosmological probe independent of the growth of structures
\cite{Jain03, Zhang05}, has been used to strongly constrain deviations from
general relativity \cite{Wang07} and has allowed a consistency test of the standard 
cosmological model \cite{Ruiz15}.

Our work extends previous studies to observables beyond the power
spectrum. In particular, we analyze the equilateral bispectrum and two
simple but promising topological observables: lensing peaks and
Minkowski functionals. We restrict our analysis to two parameters that
can influence lensing observables significantly through both geometry and growth:
the total matter density ($\Omega_m$) and the DE equation of state as parametrized
with a constant ratio of its pressure to its energy density ($w$). Future work should
include a full cosmological parameter set.
We disentangle the two contributions by measuring observables over a
collection of mock $\kappa$ maps built from ray-tracing N-body
simulations.

The paper is organized as follows. In \S~\ref{methods}, we describe
the suite of simulations we used and our method to separate the
effects of geometry and growth on the observables.
In \S~\ref{results}, we show the sensitivity of each observable to
both $\Omega_m$ and $w$, discussing the separate contributions from
geometry and growth, and in \S~\ref{inference} we show how they
impact parameter inference.  We then discuss our results in
\S~\ref{discussion} and summarize our conclusions in
\S~\ref{conclusions}.

\section{Disentangling geometry from growth in simulations}\label{methods}

We measured lensing observables on mock $\kappa$ maps generated for 9
flat $\Lambda$CDM cosmologies. We considered only DE models with a constant
ratio of pressure to energy density ($w$). Apart from $w$, we also varied $\Omega_m$, with
a fiducial model corresponding to $\{\Omega_m, w\}=\{0.26,
-1.0\}$ and the remaining 8 cosmologies each differing from it in just
one parameter (see Table~\ref{tablemodels}). For all models, we fixed the amplitude of perturbations at $\sigma_8$=0.8, the Hubble constant to $h=0.72$, the spectral index to
$n_s$=0.96 and the effective number of relativistic degrees of freedom
to $N_{\rm eff}$ = 3.04.

\begin{table}
  \caption{\label{tablemodels}\it Parameters of the eight models explored around the
    fiducial model ($\Omega_m = 0.26$, $w = -1.0$).  All models are
    spatially flat with $\Omega_{\Lambda} = 1-\Omega_m$ and consider a constant
    equation of state parameter $w$ for DE.}
  \vspace{0.5\baselineskip}
  \begin{tabular}{ccc|ccc}
\hline
\hline
$\Omega_m$ 	& $w$  & & & $\Omega_m$	& $w$  \\
\hline
0.20		&-1.0 & & & 0.26	&-0.5\\
0.23		&-1.0 & & & 0.26	&-0.8\\
0.29		&-1.0 & & & 0.26	&-1.2\\
0.32		&-1.0 & & & 0.26	&-1.5\\
\hline
\hline
\end{tabular}
\end{table}

\subsection{Simulating weak lensing maps}\label{simulations}

A set of mock convergence maps was generated by raytracing through the
outputs of dark matter-only N-body simulations, following the multiple
lens plane algorithm implemented in \textsc{Lenstools}. We used full ray-tracing to 
avoid any potential bias in the convergence descriptors under study. While
it has been shown that the Born approximation is accurate for the galaxy
lensing power spectrum \cite{Krause10} and bispectrum \cite{Dodelson05}, it can 
introduce significant biases for higher-order moments \cite{PetriBorn} and 
its effects on topological descriptors are yet unclear. We give a
brief outline of our simulation pipeline here, and refer readers for a
detailed description in \cite{Petri16LensTools}.

The observer's past light cone is discretized in a set of lens planes
separated by a constant comoving distance of $80 \, h^{-1} \rm
Mpc$. For each cosmology, we evolved the matter density field in a
single box of side $240\, h^{-1} \rm Mpc$, which can
cover a field of view of $3.5 \times 3.5 \deg^2$ up to a redshift
$z\approx$ 3. The N-body simulations were run using \textsc{Gadget2}
\cite{Springel05} with the same initial conditions. Each box contains
$512^3$ particles, yielding a mass resolution of
$\approx10^{10}M_{\odot}$. All simulation volumes were randomly
shifted and rotated to generate $1024$ different $\kappa$ maps for
each cosmology. This is justified by previous work
\cite{Petri16Variance}, which has shown that a single N-body
simulation can be recycled to generate as many as $\approx 10^{4}$
statistically independent realizations of the projected 2D convergence
field.

Bundles of $1024 \times 1024$ uniformly distributed rays were traced back to the lensed
galaxies' redshift and the convergence was reconstructed from the
accumulated deflection of the rays by the discrete lens planes. For
simplicity, we assumed all source galaxies are uniformly distributed
at a single redshifts, chosen to be either $z_s=1$ or $z_s=2$.

\begin{figure}
\begin{center}
\includegraphics[width=0.5\textwidth]{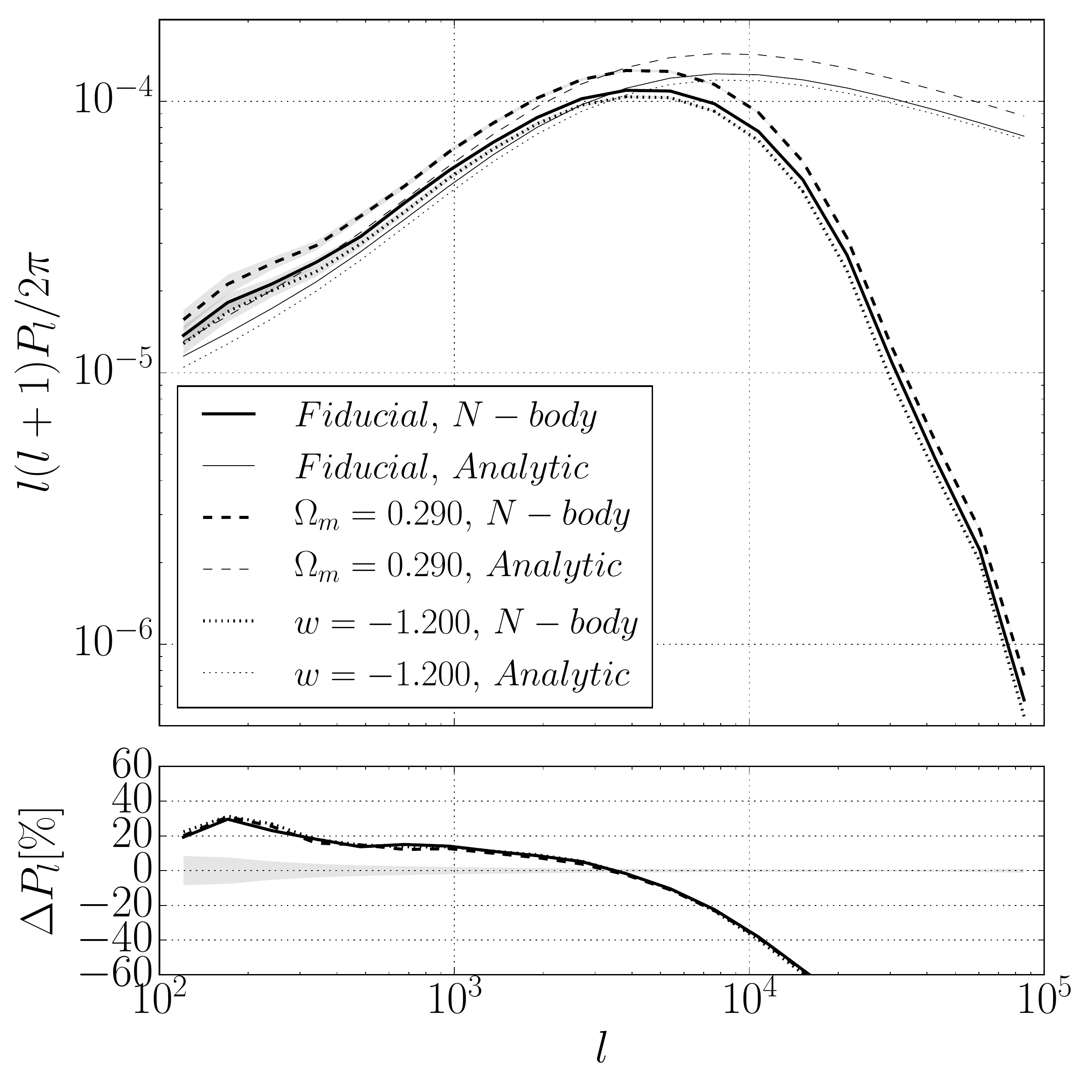}
\end{center}
\caption{\it Comparison between the power spectra measured for
  selected models, as labeled, over noiseless, un-smoothed $\kappa$
  maps (thick lines) and analytic predictions using a fitting formula
  \cite{Smith03} for the matter power spectrum (thin
  lines). Percent differences between measured and predicted power
  spectra are depicted in the lower panel. Shaded areas represent $\pm
  1$ standard deviations around the average, scaled to a $1000 \deg^2$
  survey, and in the lower panel only the standard deviation for the
  fiducial model is plotted for reference.  }\label{powerspectrum}
\end{figure}

We included the effect of galaxy shape noise assuming it is
uncorrelated with the lensing signal and its probability distribution
function (PDF) is a Gaussian with zero mean. The variance of the shape
noise depends on the r.m.s. intrinsic ellipticity noise
($\sigma_{\epsilon}$), the source galaxy surface density ($n_{gal}$)
and the pixel size ($\theta_p$), as \cite{Waerbeke00}

\begin{eqnarray}
\sigma_{p}^2 = \frac{\sigma_{\epsilon}^2}{2 n_{gal} \theta_{p}}.
\end{eqnarray}

For this work we considered an intrinsic ellipticity noise of
$\sigma_{\epsilon}=0.4$ and a galaxy density of $n_{gal}=25$
arcmin$^{-1}$, similar to the expectation for LSST but conservative
compared to the galaxy densities expected in deeper surveys, such as
Euclid and WFIRST. We generated a single set of 1024 noise-only maps
and added them to the noiseless $\kappa$ maps ray-traced from the
N-body simulations. We smoothed the noiseless $\kappa$ and shape noise
maps with the same 2D Gaussian kernel,
\begin{eqnarray}
W \left( \theta \right) = \frac{1}{2\pi\theta_S^2} \exp{\left[-\frac{\theta^2}{2\theta_S^2}\right]},
\end{eqnarray}
with $\theta$ the angular distance to each pixel, and a characteristic width 
$\theta_S =1$ arcmin. In this analysis we did not combine different smoothing
scales. The smoothing suppresses power on small scales corresponding to 
spherical multipoles on the sky $\ell \gtrapprox 12000$, which corresponds to
the scale at which we are still not limited by the finite resolution of our simulations 
(see Fig.\ref{powerspectrum}). We do not show results beyond $l=10000$, and 
the topological features, measured on the smoothed maps, do not contain
information from smaller scales.

\subsection{Isolating the effect of geometry vs. growth}\label{theory}

Galaxy shape distortions by gravitational lensing result from the
convolution of the lens properties and the distances between source
galaxies, lenses and the observer. Both effects depend on cosmology;
the former through the evolution of mass inhomogeneities, and the
latter through the expansion history of the Universe. To account for
these effects separately in our simulations, we evolved the matter
density field according to a cosmological model, but during the
ray-tracing, we allowed distances to correspond to a different
cosmology.

In our implementation of the multi-plane algorithm, lens planes are
located at the same comoving distances from the observer for all
models and we disentangled growth and geometry by 
modifying the lens planes' properties.

The lensing potential for a lens at a comoving distance of $\chi_i$, given a set of cosmological
parameters $\mathbf{p}$, is determined by its mass surface density,
\begin{eqnarray}
  \Sigma_{i}\left( x,y;\mathbf{p} \right) = \frac{3H_0^2\Delta}{2c^2} \frac{\chi_i}{a(\chi_i,\mathbf{p})}\delta\Omega_m(x,y;z(\chi_i,\mathbf{p});\mathbf{p}),
\end{eqnarray}
where $(x,y)$ are angular positions on the lens plane, $\Delta$ is the plane's thickness ($80\,h^{-1}$Mpc), $\chi$ the comoving distance, $a$ the scale factor and
$\delta\Omega_m$ the product of the density contrast and the matter
density parameter. The sensitivity of an observable to cosmology
refers to the change in that observable for a set of parameters
$\mathbf{p}$ relative to the same observable for a fiducial model
$\mathbf{p_0}$.

The effect of geometry can be estimated by evolving the perturbations
according to $\mathbf{p_0}$ and evaluating them at redshift
$z(\chi_i,\mathbf{p})$, keeping the geometrical prefactor $\chi/a$ equal
to the value that corresponds to the cosmological model
$\mathbf{p}$. Conversely, the effect of the growth of structures can
be captured by keeping the geometrical prefactor equal to its value in the fiducial model and
evaluating the density perturbations at $z(\chi_i,\mathbf{p_0})$ after evolving them according to $\mathbf{p}$.

\begin{widetext}
\begin{eqnarray}
\begin{aligned}
\Sigma_{i}^{Geometry}\left( x,y;\mathbf{p};\mathbf{p_0} \right) = \frac{3H_0^2\Delta}{2c^2} \frac{\chi_i}{a(\chi_i,\mathbf{p})}\delta\Omega_m(x,y;z(\chi_i,\mathbf{p});\mathbf{p_0})
\end{aligned}
\end{eqnarray}

\begin{eqnarray}
\begin{aligned}
\Sigma_{i}^{Growth}\left( x,y;\mathbf{p};\mathbf{p_0} \right) = \frac{3H_0^2\Delta}{2c^2} \frac{\chi_i}{a(\chi_i,\mathbf{p_0})}\delta\Omega_m(x,y;z(\chi_i,\mathbf{p_0});\mathbf{p})
\end{aligned}
\end{eqnarray}
\end{widetext}

This approach does not require running separate N-body simulations to
generate growth-only and geometry-only convergence maps, but it
involves saving additional \textsc{Gadget2} snapshots, since fixed comoving
distances correspond to different scale factors for different cosmologies.  
For each model $\mathbf{p}$, additional snapshots at redshifts
$z(\chi_i, \mathbf{p_0})$ are needed. For the fiducial cosmology, we saved additional
snapshots at redshifts $z_k(\chi_i, \mathbf{p_k})$ for each
$\mathbf{p_k}$ model considered.

\section{Sensitivity to $\Omega_m$ and $w$}\label{results}

The percentage deviation of an observable relative to its value in the
fiducial model measures its sensitivity to changes in cosmology. For
galaxy lensing, we are interested in observables measured over
$\kappa$ maps that include shape noise. We focus on the behavior of
four observables: the power spectrum, which has already been studied
analytically and will serve as a test of our simulation-based
approach, the equilateral bispectrum, which should be zero for a
Gaussian random field, and two topological features that have been
used to probe non-Gaussianities: lensing peaks and Minkowski
functionals. We measured the sensitivities from the full ray-traced
N-body simulations, as well as from simulations that only capture the
changes due to either the expansion history or to the structure growth
in a given cosmology.

\subsection{Power spectrum}\label{ps}

The convergence power spectrum is the Fourier transform of the 2-point
correlation function of $\kappa(x,y)$ and is one of the most popular
weak lensing observables. For a flat cosmology, with lensed sources at a fixed
redshift, and using the Limber and flat-sky approximations,
the power spectrum can be expressed as a line-of-sight integral of
the matter power spectrum, weighted by a geometrical kernel
\cite{Schneider98}

\begin{eqnarray}\label{powerspectrumequation}
\begin{aligned}
P_k(l) &= \frac{9}{4}\left(\frac{H_0}{c}\right)^4\Omega_m^2\int_0^{\chi_s} \frac{d\chi}{a^2(\chi)} \left(1-\frac{\chi}{\chi_s}\right)^2 P_\delta \left( \frac{l}{\chi}; \chi \right)
\end{aligned}
\end{eqnarray}

Where $\chi$ is the comoving distance and $\chi_s$ the  
comoving distance to the lensed galaxies. Geometry
affects the power spectrum through $\chi$ and the scale factor
$a$. Growth enters the above expression through the matter
power spectrum, $P_\delta$ (including non-linear effects), and the 
$\Omega_m^2$ outside of the integral. For our analytic calculations,
we used the \textsc{Nicaea} implementation of the convergence 
power spectrum with the prescription from \cite{Smith03} for the matter
power spectrum.

\begin{figure*}
\begin{center}
\includegraphics[width=1.0\textwidth]{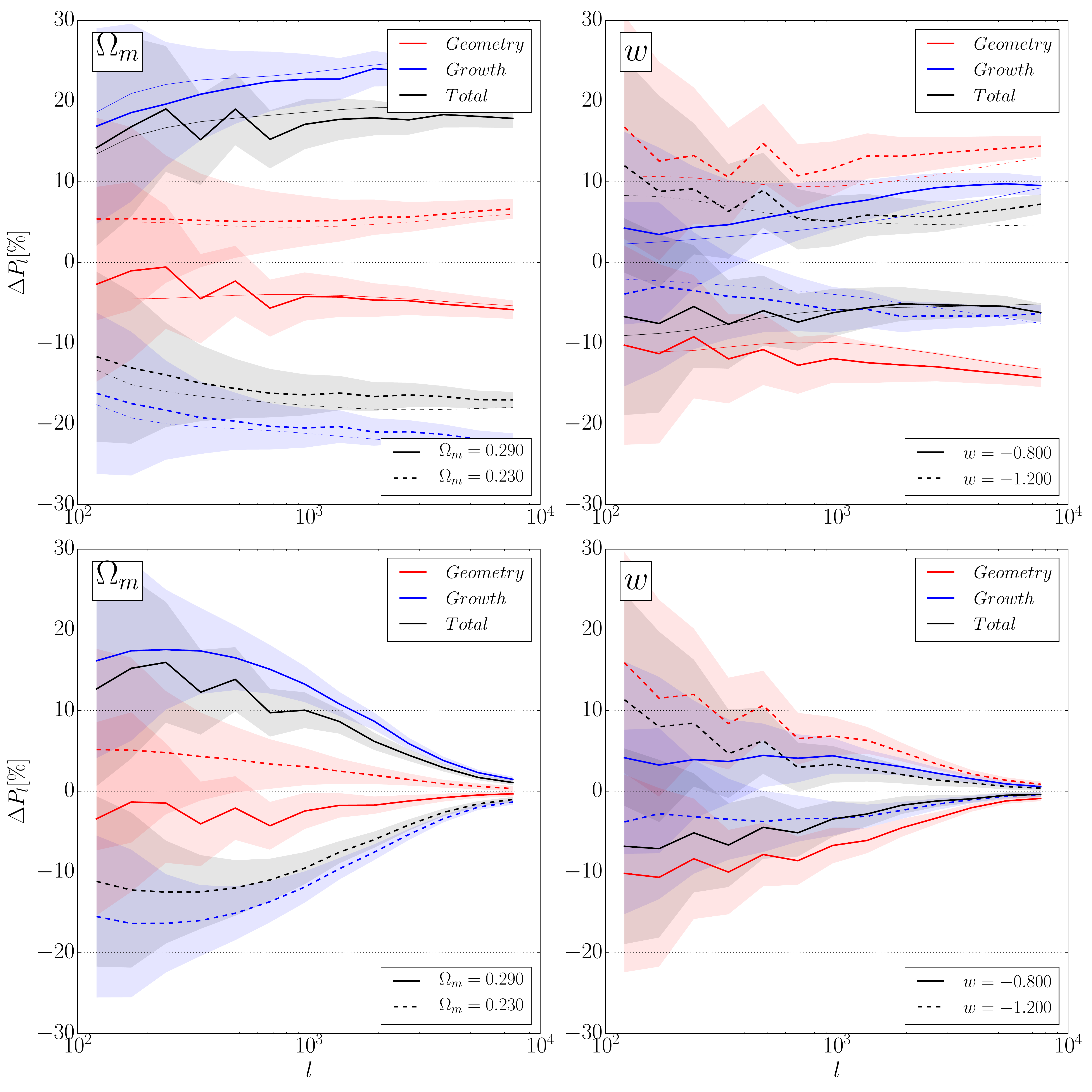}
\end{center}
\caption{ (color online)\it Sensitivity of the power spectrum to
  $\Omega_m$ and $w$ for noiseless (upper panels) and noisy (lower
  panels) convergence. Estimates including only geometry effects are
  shown in red, those including only growth effects in blue, and those
  including both effects in black. In the upper panels, analytic
  predictions are displayed with thin lines, for comparison. Source
  galaxies are at $z_s=1.0$ in all cases. Shaded areas represent a
  $\pm 1$ standard deviation around the measured averages scaled to a
  survey sky coverage of 1000 $\deg^2$ and only selected models are
  displayed for clarity.  } \label{splitps}
\end{figure*}

We determined the percentage deviation of the power spectrum relative
to the fiducial cosmology over 1024 noiseless, un-smoothed $\kappa$
maps for each non-fiducial cosmology, and compared the results with
analytic predictions.  These results, shown in the upper panels of
Fig.~\ref{splitps}, match the analytic predictions within the
statistical uncertainties, and are also in good agreement with the
findings of \cite{Simpson05}. The sensitivity is only weakly dependent
on the multipole.

The sensitivity to $\Omega_m$ is dominated by growth, with a $\approx 25\%$ 
change that is what would be expected from the $\approx 12\%$ change in 
$\Omega_m$ ($\Omega_m^2$ outside the integral in Eq.~\ref{powerspectrumequation}). Geometry
acts in the opposite direction, reducing the overall sensitivity by $\approx 20\%$. The
sensitivity to $w$ is dominated by geometry. While we expected its sensitivity to be smaller than that
to $\Omega_m$ due to the integrating effect, the partial cancellation 
between growth and geometry is even more severe. It reduces the sensitivity further ($\approx 50\%$) to a level of $\approx 5\%$ for a $20\%$ change in the parameter. The smaller 
sensitivity should propagate into tighter constrains on $\Omega_m$
than on $w$ from weak lensing data.

The origin of the partial cancelation is explained in detail in
\cite{Simpson05}, but we reproduce the argument here for
convenience. Making $w$ more negative, from the fiducial $w=-1.0$ to
-1.2, yields a higher DE density in the past. The comoving distance to
the source galaxies' redshift becomes larger, and so does the
cumulative effect of small deflections experienced by light rays. As a
result, the effect due to geometry is an increase of the lensing
signal. Since we fix the amplitude of the perturbations at the present
time ($\sigma_8$) in our simulations, a higher DE density in the past
means there are fewer structures to deflect the light rays in the
past, and the growth contribution to the lensing signal is smaller 
compared to a model with constant dark matter density.

Galaxy shape noise introduces a scale-dependence to the relative
sensitivity, as clearly seen in the lower panels of
Fig.\ref{splitps}. At small scales, white noise dominates the power
spectrum and suppresses its sensitivity to cosmological
parameters. Galaxy shape noise then limits the information that can be
extracted from the convergence power spectrum at small scales.

\subsection{Equilateral bispectrum}\label{bs}

The natural extension to the two-point correlation function is the
three-point correlation function, or its Fourier transform, the
bispectrum. A non-zero bispectrum is a clear non-Gaussian signal and
has been detected in shear data \cite{Bernardeau02, Fu14}. The analog
of Eq.~(\ref{powerspectrumequation}) links the convergence bispectrum
to the bispectrum of the underlying matter density field through a
Limber integration \cite{Schneider98}

\begin{eqnarray}
\begin{split}
\mathbf{B}_k\left(\boldsymbol{l}_1,\boldsymbol{l}_2,\boldsymbol{l}_3\right) = \frac{27}{8}\left(\frac{H_0}{c}\right)^6\Omega_m^3 \int_0^{\chi_s} \frac{d \chi}{\left( \chi a(\chi) \right)^3} \\ 
\left( 1-\frac{\chi}{\chi_s} \right)^3  \delta^D(\boldsymbol{l_1} + \boldsymbol{l_2} + \boldsymbol{l_3}) \mathbf{B}_{\delta}\left( \frac{\boldsymbol{l}_1}{\chi}, \frac{\boldsymbol{l}_2}{\chi}, \frac{\boldsymbol{l}_3}{\chi}; \chi \right)
\end{split}
\end{eqnarray}

\begin{figure*}
\begin{center}
\includegraphics[width=1.0\textwidth]{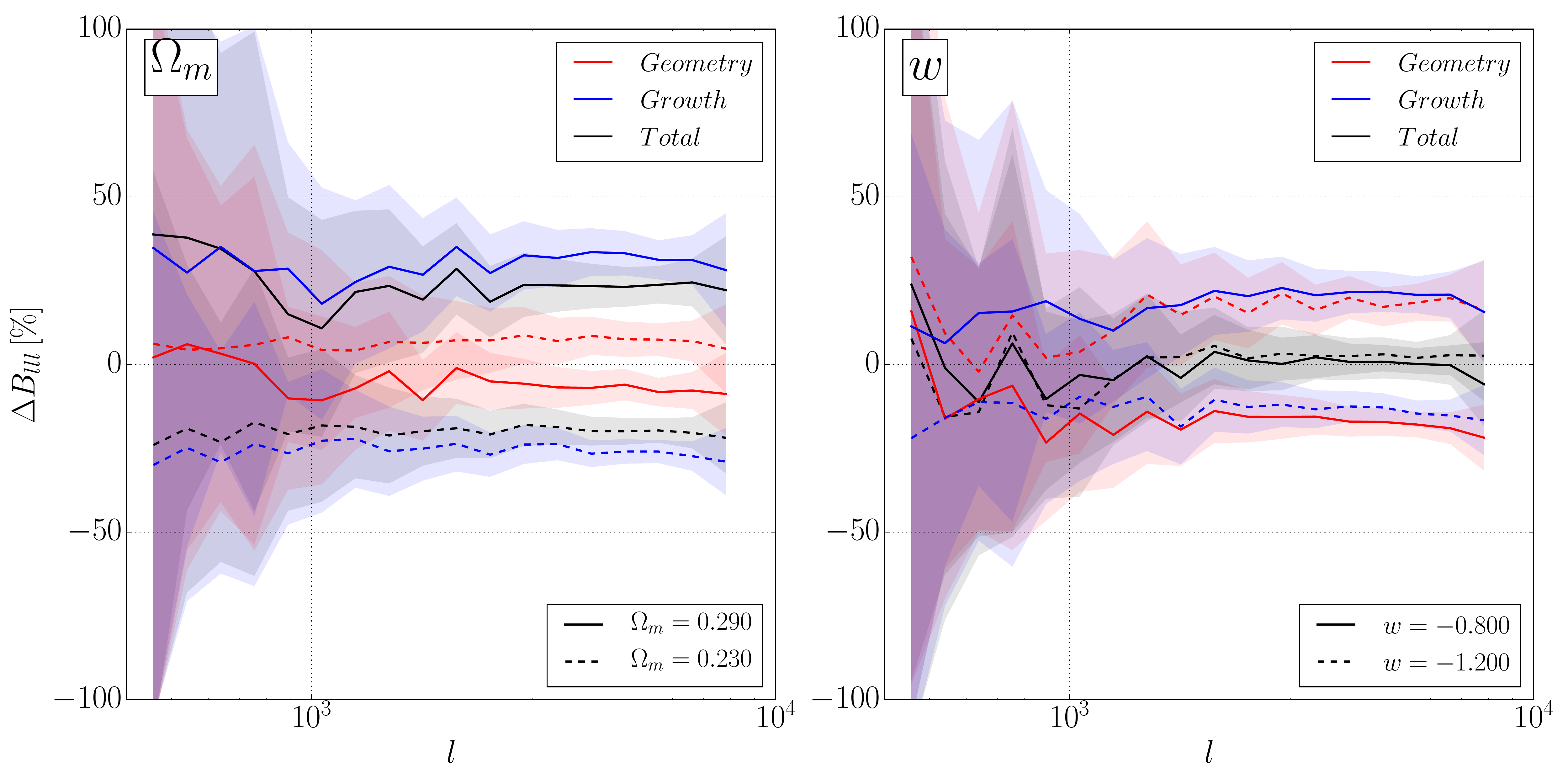}
\end{center}
\caption{ (color online)\it Sensitivity of the equilateral bispectrum of
  the noiseless convergence field to $\Omega_m$ and $w$. Both panels
  show the percentage deviation in each model from the fiducial
  bispectrum. For clarity, only two models are depicted per panel,
  with the source galaxies at $z_s=1$. As in Fig.~\ref{splitps}, black
  lines show the net sensitivity, red lines the sensitivity due only
  to differences in geometry and blue lines the sensitivity due only
  to differences in growth. Shaded areas represent $\pm 1$ standard
  deviation around the measured averages, scaled to a 1000 $\deg^2$
  survey.  } \label{splitbispectrum}
\end{figure*}

Where $\delta^D$ is a Dirac delta. When the lengths of the triangle defined by the three points on which
the correlation function are measured are the same, the result is the
equilateral bispectrum ($\mathbf{B}_{lll}$). In an exercise analogous
to the one done for the power spectrum, we measured $\mathbf{B}_{lll}$
for our mock noiseless convergence maps and show their relative
sensitivity to the cosmological parameters in
Fig.~\ref{splitbispectrum}.

While noisier, the parameter-sensitivity has a behavior very similar
to the case of the power spectrum, in terms of its weak dependence on
the angular scale $\ell$, order of magnitude, and split between geometry
and growth. The most noticeable difference is that the cancelation
between both effects is almost perfect for $w$, resulting in a
statistic that is almost insensitive to that parameter. The results
for the lensed galaxies at $z_s=2$ are similar, and show the same
cancelation for $w$. The addition of shape noise results in an even
noisier measurement (see \S\ref{inference}) with error bars 3-4 times
larger than the ones displayed in Fig.~\ref{splitbispectrum} for the
noiseless case. There is no average sensitivity suppression at small scales,
because the shape noise is Gaussian.

\subsection{Lensing peaks}\label{peaks}

Peaks, defined as local maxima on smoothed $\kappa$ maps, probe
high-density regions, where non-Gaussianities of the convergence
should be enhanced. Also, they are computationally inexpensive to
measure, making them an attractive observable to combine with others
for cosmological inference. Indeed, their distribution as a function
of their height, or peak function, has been forecast to improve
constraints obtained using only second-order statistics by a factor of
$2-3$ \cite{Kratochvil10, Dietrich10}. Similar improvements have
now been found in recent lensing survey data \cite{Liu15, Liu15NotJia, DES16}.

\begin{figure*}
\begin{center}
\includegraphics[width=1.0\textwidth]{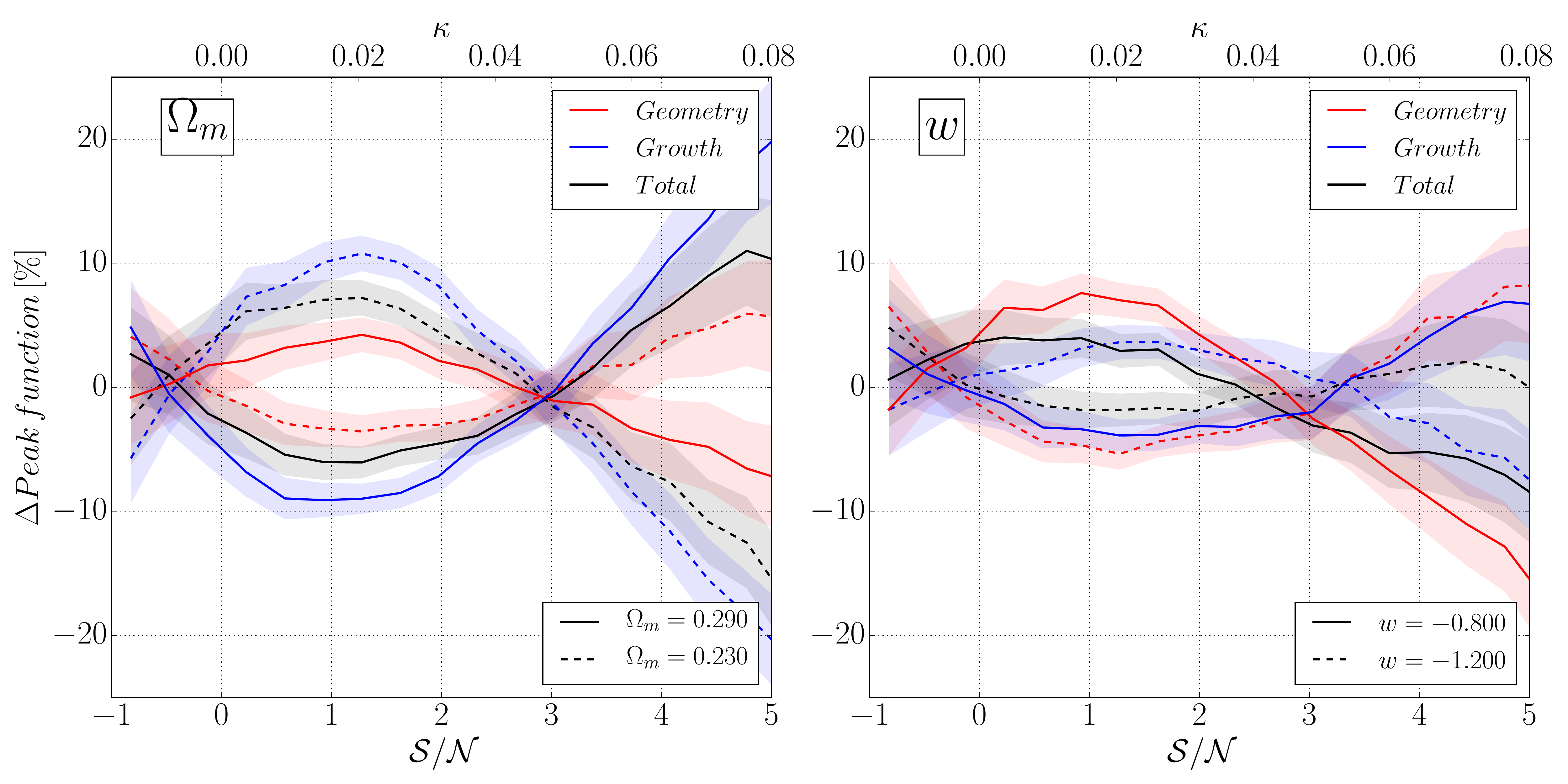}
\end{center}
\caption{(color online)\it Sensitivity of peak counts to $\Omega_m$ and
  $w$ on noisy convergence maps. Both panels show the percentage
  difference between the peak counts in a given cosmology and in the
  fiducial model. Peak height is expressed in units of $\kappa$ and in units of $\sigma_{noise}$, $\mathcal{S/N}$. For clarity, only two models are depicted per panel,
  with source galaxies at $z_s=1$.  The color scheme is the same as in
  Figs.~\ref{splitps}~and~\ref{splitbispectrum}.  Shaded areas
  represent $\pm 1$ standard deviation around the measured averages,
  scaled to a 1000 $\deg^2$ survey.  } \label{splitpeaks}
\end{figure*}

We extracted peak catalogues from our mock convergence maps and
computed the percentage deviation of the peak height function relative
to the fiducial model. The results for the noisy case are shown in
Fig.\ref{splitpeaks}.  We again observe some similarities between the
sensitivity of the peak height functions and that of the power
spectrum. The $\Omega_m$-sensitivity is dominated by growth, while
geometry dominates the sensitivity to $w$. There is also a partial
cancelation between the two effects, and the cancelation is stronger
for $w$, yielding a reduced net sensitivity compared to $\Omega_m$, by
a factor of $\approx 2$.

For high peaks, the sign of the parameter-sensitivity is the same as
for the power spectrum, but the sign reverses for low peaks, whose
abundance is anti-correlated with those of high peaks. High peaks are
$\approx 2-3$ times more sensitive than low peaks, but there are fewer
of them to help discern between models (see \S\ref{discussion}). Shape noise 
modifies the peak function, by introducing new peaks, eliminating some, and 
spreading the height of those that survive from the noiseless maps. As
a result, it reduces the sensitivity by a factor of
$\approx 2$, especially for the noise-dominated low peaks, and moves
the turn-over point, where the parameter-sensitivity changes sign,
from $\mathcal{S/N} \approx 1$ for noiseless $\kappa$ to
$\mathcal{S/N} \approx 2.5$ ($\mathcal{S/N}$ is the height of the peaks expressed
in units of $\sigma_{noise}$).

For noisy $\kappa$ and lensed galaxies at $z_s=2$, the turn-over point
moves to even higher $\kappa$, from $\mathcal{S/N} \approx 2.5$ to $\approx 3$, and the relative
sensitivity of low peaks increases by a factor of $\approx 2$, while
the sensitivity of high peaks remains the same.

\subsection{Minkowski functionals}\label{minkowski}

Minkowski functionals (MFs) on 2D fields are topological measures on
iso-contours \cite{Mecke94}. They capture statistical information of
all orders and have been shown to constrain cosmology, improving
errors computed exclusively from the power spectrum, in theoretical
studies \cite{Kratochvil12} and also when applied to observations
\cite{Shirasaki14, Petri15}.

\begin{figure*}
\begin{center}
\includegraphics[width=1.0\textwidth]{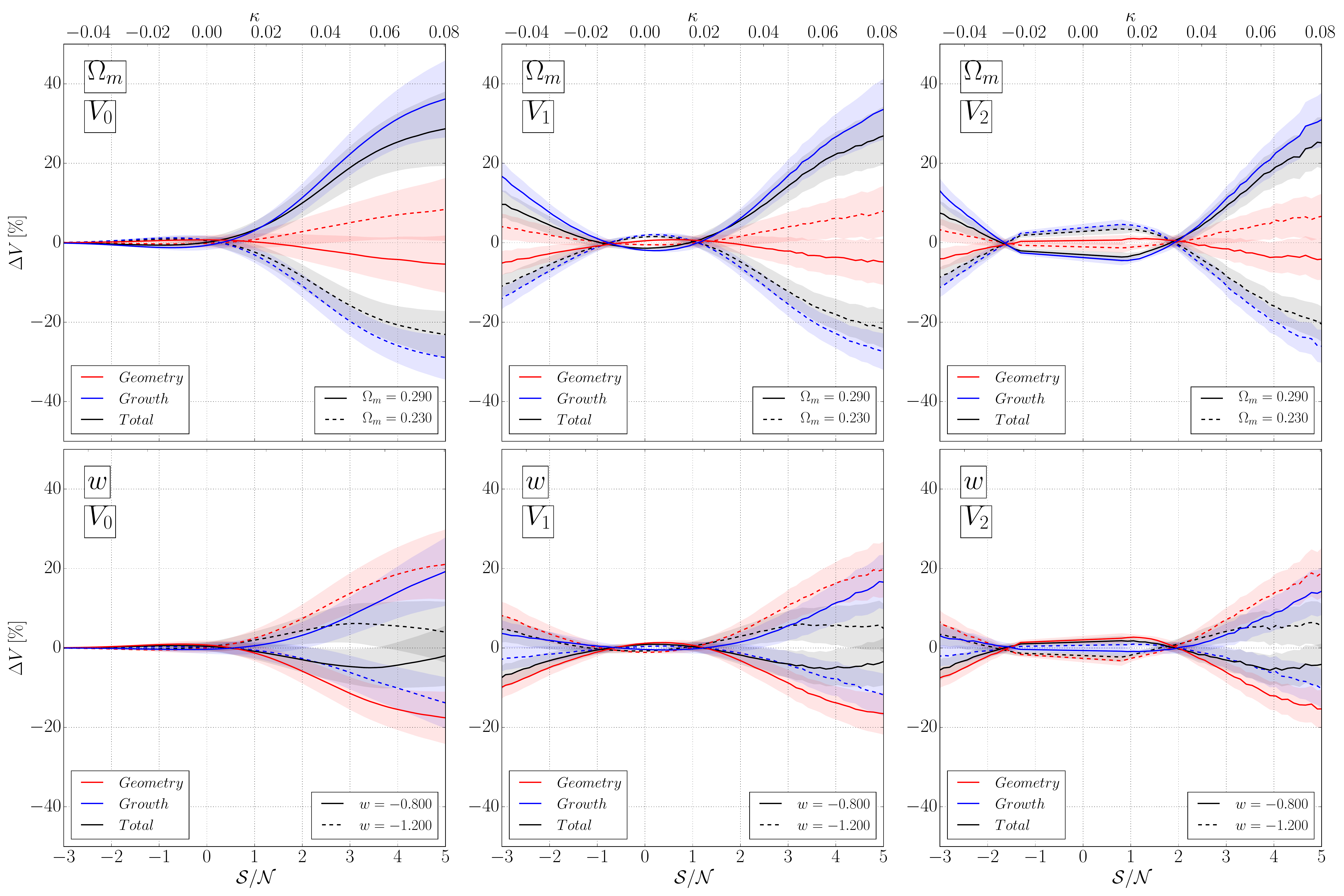}
\end{center}
\caption{(color online)\it Percentage difference of the three MFs
  measured on noisy $\kappa$ maps, compared to the value in the
  fiducial model, when changing $\Omega_m$ and $w$. Left/center/right
  panels show the results for $V_0$/$V_1$/$V_2$, for noisy $\kappa$
  and source galaxies at $z_s=1$. The color scheme, labeled in the
  legends, is the same as in Figs.~\ref{splitps}-\ref{splitpeaks}.
  Shaded areas represent $\pm 1$ standard deviation around the
  measured averages, scaled to a 1000 $\deg^2$ survey.
} \label{fig:minkowski}
\end{figure*}

The three MFs on a 2D map measure the area ($V_0$), boundary length
($V_1$) and the Euler characteristic ($V_2$) of the set of points
where the value of the function exceeds a pre-specified threshold
($\kappa_{\rm th}$):
\begin{eqnarray}
V_0(\kappa_{\rm th}) = \frac{1}{A} \int_{\Sigma(\kappa_{\rm th})} da, \\
V_1(\kappa_{\rm th}) = \frac{1}{4A} \int_{\partial \Sigma(\kappa_{\rm th})} dl, \\
V_2(\kappa_{\rm th}) = \frac{1}{2\pi A}\int_{\partial \Sigma(\kappa_{\rm th})} \kappa dl,
\end{eqnarray}
where $A$ is the total area of the map, $\Sigma(\kappa_{\rm th})$ is
the set of points on the convergence map for which
$\kappa\geq\kappa_{\rm th}$, and $\partial\Sigma(\kappa_{\rm th})$
denotes a line integral along the curve where $\kappa=\kappa_{\rm th}$.
We refer the reader to \cite{Kratochvil12} for a detailed description
of our measurement procedure, and reproduce in
Fig.~\ref{fig:minkowski} the percentage difference between the MFs for
a given cosmology and the fiducial model, as a function of the
threshold.

The sensitivity of all three functionals at high threshold levels is
similar to that of peak counts. This is expected, since at high
$\kappa_{\rm th}$ values, the set of points $\kappa\geq\kappa_{\rm
  th}$ increasingly coincides with the set of lensing peaks. At lower
thresholds, the sensitivity of the MF is weaker, but different for
each functional, suggesting that combining them should yield tighter
parameter constraints.

\section{Impact on parameter inference}\label{inference}

Parameter constraints are not just determined by the sensitivity of
observables, but also by their (co)variances. To assess the impact of
geometry and growth on inference, we estimated the confidence levels
on the parameters $(\Omega_m, w)$ in two ways. First, we quantified
how different each model is from the fiducial, using the $\Delta\chi^2$,
\begin{eqnarray}
\Delta \chi^2 = \sum_{i,j} \left( \mu_i - \mu_i^{fid} \right) C^{-1}_{ij} \left( \mu_j - \mu_j^{fid} \right),
\end{eqnarray}
where $\mu_i$ is the average of an observable over the set of
convergence maps for a cosmology (for instance, the binned power spectrum), 
$\mu_i^{fid}$ the average for the fiducial cosmology and $C^{-1}_{ij}$ the precision matrix. 
For each observable we used 20 bins, either spaced logarithmically in $\ell$ or
linearly in $\kappa$. We did not try to optimize the number of bins or
their thresholds, since our purpose was to understand the effect of
geometry and growth on the parameter uncertainties, not obtain
accurate or optimal estimates for a specific survey.

We computed the precision matrix in the fiducial model, to be
consistent with our calculated Fisher matrices (see below), and we
corrected for its bias following \cite{Hartlap07}. The bias correction
is very small, $\approx 2\%$, because the number of realizations used
to estimate the covariance matrix ($N_r=1024$) is large compared to
the dimensionality of the data vector ($N_b=20$). We scaled the
results by the same factor as the error bars in the figures, so that
their magnitude corresponds to what would be expected for a 1000
$\deg^2$ survey, even though in the non-Gaussian regime errors may
scale logarithmically rather than as the square root of the field of
view \cite{Creminelli07}.

\begin{table*}
\caption{\label{chi2} \it $\Delta \chi^2$ for different cosmological
  models computed for the power spectrum and three non-Gaussian
  observables (equilateral bispectrum, peak counts and Minkowski
  functionals) over noisy $\kappa$ maps with source galaxies at either
  $z=1$ or $z=2$. }
\begin{ruledtabular}
\begin{tabular}{lcccccccccccccccc}
&
\multicolumn{8}{c}{Dependence on $\Omega_m$}  &
\multicolumn{8}{c}{Dependence on $w$} \\

\cline{2-9}
\cline{10-17}

& 
\multicolumn{2}{c}{0.200} &
\multicolumn{2}{c}{0.230} &
\multicolumn{2}{c}{0.290} &
\multicolumn{2}{c}{0.320} &

\multicolumn{2}{c}{-0.500} &
\multicolumn{2}{c}{-0.800} &
\multicolumn{2}{c}{-1.200} &
\multicolumn{2}{c}{-1.500} \\

\cline{2-3}
\cline{4-5}
\cline{6-7}
\cline{8-9}
\cline{10-11}
\cline{12-13}
\cline{14-15}
\cline{16-17}

			&$z=1$ &$z=2$ &$z=1$ &$z=2$ &$z=1$ &$z=2$ &$z=1$ &$z=2$ &$z=1$ &$z=2$ &$z=1$ &$z=2$ &$z=1$ &$z=2$ &$z=1$ &$z=2$ \\
\hline
& \multicolumn{16}{c} {Power spectrum} \\
\hline
Total			&541&1550  &148&421  &174&444  &718&1770  &71&288  &18&45  &22&18  &109&109 \\
Geometry-only	&92&670  &18&142  &14&101  &48&375  &525&2442  &92&379  &102&271  &532&1569  \\ 
Growth-only	&839&3033 &242&861  &305&1110  &1371&5083 &528&3050  &45&252  &23&132  &96&557       \\
\hline
& \multicolumn{16}{c} {Equilateral bispectrum} \\
\hline
Total			&14&38  &3&13  &8&8     &25&41   &4&5      &2&3    &3&4   &3&5       \\
Geometry-only	&5&9      &2&5    &2&4     &2&10     &14&47  &4&12  &7&8   &19&40       \\
Growth-only	&18&56  &6&16  &12&20 &40&113 &39&181 &4&13  &3&8   &9&28       \\
\hline
& \multicolumn{16}{c} {Peak counts} \\
\hline
Total			&772&1120        &190&266        &199&232        &768&825       &211&399            &39&48         &38&26           &164&93       \\
Geometry-only	&99&336          &26&76          &23&70          &65&223         &776&1934            &127&253       &110&178         &603&837       \\
Growth-only	&1213&2431      &317&588        &361&542        &1445&2071     &321&931           &40&114         &20&83            &117&373       \\
\hline
& \multicolumn{16}{c} {Minkowski functional $V_0$} \\
\hline
Total			&915&1153        &231&272        &212&265         &859&976        &413&828          &64&90          &52&56          &268&194       \\
Geometry-only	&111&455        &28&107          &30&81           &86&281          &931&2305          &150&282        &126&229        &711&1071       \\
Growth-only	&1464&2684      &386&650        &404&651         &1663&2634      &385&1189          &38&121          &26&75          &116&337       \\
\hline
& \multicolumn{16}{c} {Minkowski functional $V_1$} \\
\hline
Total			&984&1506       &245&339       &229&353       &901&1229       &321&516            &52&53         &41&33           &205&112       \\
Geometry-only	&118&422    &27&117         &29&73         &88&271         &996&2595          &161&285       &130&214         &696&1075       \\
Growth-only	&1564&3313       &400&799       &423&753       &1691&3043       &635&2068            &61&199         &34&119           &158&543       \\
\hline
& \multicolumn{16}{c} {Minkowski functional $V_2$} \\
\hline
Total			&1016&1862       &255&438       &253&446        &997&1647      &313&486          &56&51           &39&34         &203&109       \\
Geometry-only	&128&602         &30&141         &31&101          &95&375        &1030&3206        &173&392        &145&292        &764&1412      \\
Growth-only	&1613&4068       &420&1000      &460&997        &1910&4031      &736&2832          &68&280           &39&157         &164&712       \\
\end{tabular}
\end{ruledtabular}
\end{table*}

The $\Delta \chi^2$ values are listed in Table~\ref{chi2}, and are
overall consistent with the conclusions from the sensitivity plots in
\S~\ref{results}. The significance at which models with different
$w$'s can be distinguished is lower than for $\Omega_m$, due to 
projection effects and the worse cancelation between geometry and growth. 
Geometry has stronger constraining power in $w$ and growth does
in $\Omega_m$, and in general the net significance is closer to that
of growth than that of geometry. The observable with the lowest
$\Delta \chi^2$ is the equilateral bispectrum, especially for $w$, for
which the cancelation between geometry and growth is particularly
severe.

Even though it can strictly be used only for Gaussian-distributed data, we computed the Fisher matrix 
\cite{Tegmark97} for all the observables in this study, with the expectation that it provides a
second-order approximation to the true  parameter likelihood near its maximum:
\begin{eqnarray}
\begin{split}
F_{\alpha \beta} &= \frac{1}{2} {\rm Tr} \left[ C^{-1}C_{,\alpha} C^{-1} C_{,\beta} + C^{-1} M_{\alpha\beta}\right],\\
M_{\alpha \beta} &= \mu_{,\alpha} \mu_{,\beta}^T + \mu_{,\beta} \mu_{,\alpha}^T.
\end{split}
\end{eqnarray}
Here $F_{\alpha \beta}$ is one element of the Fisher matrix, ${\rm
  Tr}$ stands for the trace of the matrix within brackets, the
covariance is evaluated at the fiducial model and a comma denotes them
partial derivative $X_{,\alpha}\equiv \frac{\partial}{\partial
  \alpha}X$. The marginalized error on a parameter is given by
$\sqrt{\left( F^{-1} \right)_{\alpha \alpha}}$, and is reported in Table~\ref{Ferrors}.
We have found the finite-difference derivatives of
the covariance to be sensitive to the numeric scheme used to estimate
them, especially for the bispectrum. In the case of the power spectrum and peak counts,
it has been shown that this does not significantly change the parameter
constraints \cite{LKII,Zorrilla16}. For these reasons, we have not
included the cosmology-dependence of the covariance in our Fisher
matrix calculations. The derivatives of the average observables were
estimated using 5-point finite differences with Lagrangian
polynomials.

\begin{table*}
\caption{\label{Ferrors}\it Marginalized errors on $\Omega_m$ and $w$,
  orientation of the Fisher ellipse (measured as the angle between its
  major axis and the $w$ axis), and figure-of-merit (FOM; defined as $\pi/A$, with $A$ the area of
  the error ellipse).  The errors correspond to a $68\%$ confidence
  level, scaled to a 1000 $\deg^2$ survey. All calculations were done
  on noisy $\kappa$ maps with source galaxies at either $z=1$ or $z=2$.}
\begin{ruledtabular}
\begin{tabular}{l  c c c c c c c c }
&
\multicolumn{2}{c}{$\Delta \Omega_m $} &
\multicolumn{2}{c}{$\Delta w$}                 &
\multicolumn{2}{c}{$\theta$}                     &\\
&
\multicolumn{2}{c}{$\times 10^{3}$} &
\multicolumn{2}{c}{$\times 10^{3}$} &
\multicolumn{2}{c}{$[\deg]$}    &
\multicolumn{2}{c}{$FOM$}    \\
\cline{2-3}
\cline{4-5}
\cline{6-7}
\cline{8-9}

			&$z=1$ &$z=2$ &$z=1$ &$z=2$ &$z=1$ &$z=2$ &$z=1$ &$z=2$ \\

\hline
& \multicolumn{8}{c} {Power spectrum} \\
\hline
Total				&14.2&4.9		&269.2&127.1       &-2.9&-2.0        &1034&3609\\
Geometry-only		&40.7&34.4	&106.6&138.3       &20.3&13.9       &802&1714\\
Growth-only		&15.2&11.6	&298.8&181.5       &2.9&3.6          &1211&3776\\
\hline
& \multicolumn{8}{c} {Equilateral bispectrum} \\
\hline
Total				&22.3&17.0	&347.2&258.4    &-0.6&1.2      	&131&241\\
Geometry-only		&49.4&38.2	&161.1&152.9    &5.9&10.2        &132&239\\
Growth-only		&34.7&34.6	&396.7&326.5    &4.3&5.7            &142&272\\
\hline
& \multicolumn{8}{c} {Peak counts} \\
\hline
Total				&8.9&7.3		&135.9&135.9  	&-3.5&-2.8       	&2247&2538\\
Geometry-only		&32.9&32.9	&98.3&128.5      &17.9&14.2       	&1087&1447\\
Growth-only		&9.4&9.8		&219.2&158.3     &2.4&3.5           &1844&3287\\
\hline
& \multicolumn{8}{c} {Minkowski functional $V_0$} \\
\hline
Total				&7.6&5.1		&99.8&66.4         &-4.0&-3.6       &3259&5387\\
Geometry-only		&29.4&36.5	&89.3&146.6       &17.6&13.9      &1311&1425\\
Growth-only		&4.8&4.5		&115.2&79.3      &2.1&3.0        &3780&7018\\
\hline
& \multicolumn{8}{c} {Minkowski functional $V_1$} \\
\hline
Total				&6.1&3.2		&91.0&63.9       &-3.3&-1.8       &3697&6355\\
Geometry-only		&38.0&36.6	&111.6&152.5       &18.4&13.4      &1042&1384\\
Growth-only		&5.1&5.6		&104.1&84.1       &2.5&3.6        &4277&7229\\
\hline
& \multicolumn{8}{c} {Minkowski functional $V_2$} \\
\hline
Total				&6.5&3.1 		&101.2&69.9       &-3.3&-1.8       &3436&6579\\
Geometry-only		&36.7&40.4	&106.5&159.0       &18.7&14.2      &1130&1489\\
Growth-only		&5.5&6.8		&109.8&99.2       &2.6&3.8       &4181&6962\\
\end{tabular}
\end{ruledtabular}
\end{table*}

\begin{figure*}
\begin{center}
\includegraphics[width=1.0\textwidth]{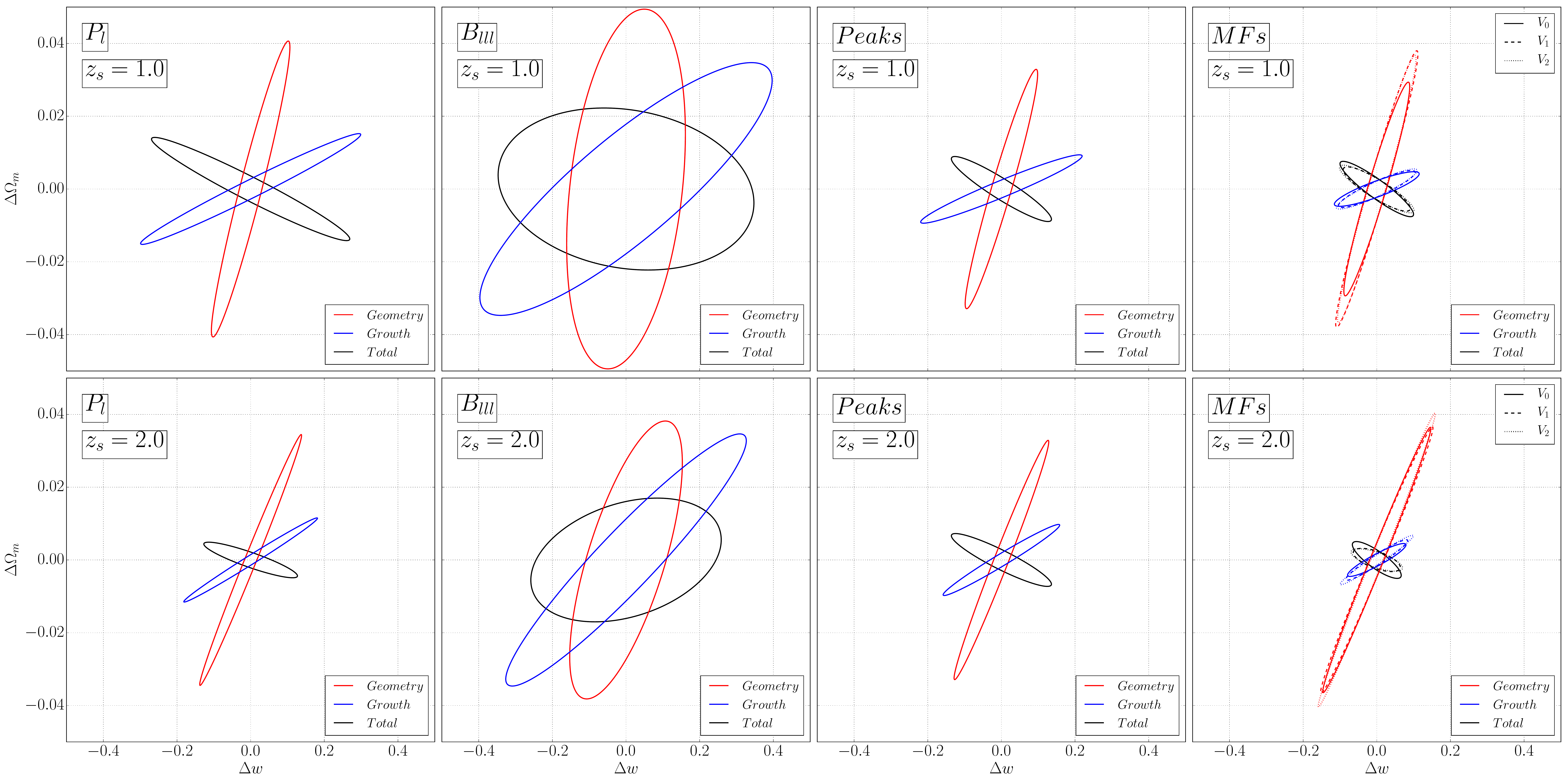}
\end{center}
\caption{(color online)\it $68\%$ Fisher error ellipses in the
  (${\Omega_m, w}$) plane inferred from the power spectrum ($P_l$),
  equilateral bispectrum ($B_{lll}$), lensing peaks and Minkowski
  functionals (MFs). Upper/lower panels show the contours for source
  galaxies at $z_s=1$/$z_s=2$. Each observable was characterized by a
  data vector of length 20, and the ellipses were computed neglecting
  the cosmology-dependence of the covariance matrix. All contours are
  scaled to a 1000 $\deg^2$ survey.  } \label{fisher}
\end{figure*}

We show the $68\%$ confidence level contours in Fig.\ref{fisher}. The
figures show that marginalized errors on $w$ are larger than those
for $\Omega_m$ by a factor of $\approx 15$, and that geometry has less
constraining power than growth. The confidence regions decrease when
the sources are farther away, although the marginalized errors do not
always do. This is due to changes in the degeneracies (i.e. the axes
and tilt angles of the error ellipses). For example, the $68\%$
contour from Minkowski functionals for geometry-only becomes more
elongated and its tilt is increased towards the $w$ axis, yielding a
larger marginalized error on $w$ for $z_s=2$ than for $z_s=1$.

For all observables, errors on $\Omega_m$ and $w$ are positively
correlated, when either geometry or growth is considered in
isolation. For example, the geometry effect of a higher matter density
is smaller comoving distances, which can also be achieved with a
less negative value for $w$. The effect on growth of a lower DE density
in the past would be a smaller suppression of gravitational collapse and a stronger
gravitational field for the collapsing perturbations. The correspondingly
stronger lensing signal is similar to what would be achieved with higher matter
density. For the net effect, the change of the dominant effect for
$\Omega_m$ and $w$ reverses the degeneracy direction, yielding
anti-correlated errors on the parameters.

\section{Discussion}\label{discussion}

The agreement between the sensitivity to $\Omega_m$ and $w$ of the
power spectra measured on the mock $\kappa$ maps and the analytic
prediction, as well as the relative contribution of geometry and
growth, validates our approach based on modified simulations.

The cancelation between geometry and growth, which further suppresses 
the sensitivity of WL to cosmological parameters, highlights why it
is important to combine different redshift bins (tomography) to
constrain DE with better precision (e.g. \cite{Hu99}). The suppression
of the power spectrum sensitivity at small scales by galaxy shape
noise highlights the importance of including other observables when
analyzing weak lensing data, even if non-Gaussianities were small.

The sensitivity of the equilateral bispectrum follows a similar
pattern to that of the power spectrum, but their measurement is
considerably noisier, which translates into a less significant $\Delta
\chi^2$ for a given model. The addition of shape noise does not affect
the mean sensitivity on small scales more than large scales, 
which is reasonable given the Gaussian noise model used (it does 
contribute to the statistical error).

We measured also the folded bispectrum, and the results are in line with
those from the equilateral shape. We expect the same for all other configurations 
of the bispectrum, for the percentage change of the power 
spectrum and bispectrum does not depend on the multipole, and the cancelation 
between geometry and growth is a feature present at map level (see below).

\begin{figure}
\begin{center}
\includegraphics[width=0.5\textwidth]{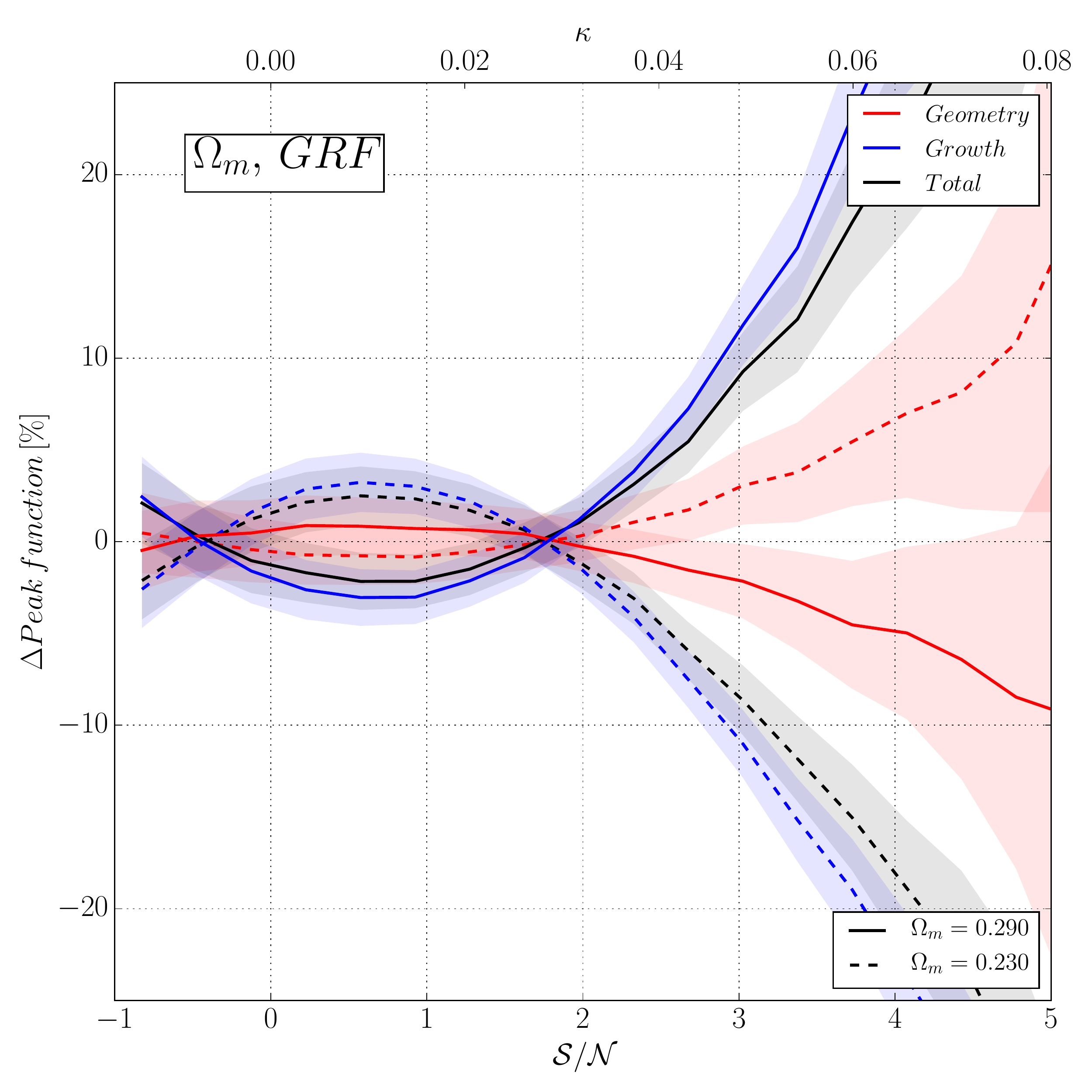}
\end{center}
\caption{(color online)\it Sensitivity of lensing peak counts to
  $\Omega_m$, derived from a set of Gaussian random fields with the
  same power spectra as that measured on noisy convergence maps from
  large-scale structure. Shaded areas represent 1 standard deviation
  errors in a 1000 $\deg^2$ survey. Compare with the left panel of
  Fig.\ref{splitpeaks} } \label{grf}
\end{figure}

The sensitivity of lensing peaks also has qualitative similarities to
that of the power spectrum, but it is highly dependent on the height
of the peaks. In order to assess how much of their sensitivity is a
direct result of differences in the power spectrum, we computed it
from Gaussian random fields (GRFs) built with the same power spectra
as the $\kappa$ maps generated through ray-tracing. The result of this
exercise is shown in Fig.~\ref{grf}. We have found that the
sensitivity of low peaks is reduced by a factor of $\approx 2$, and
the sensitivity of the high peaks increases (although there are
fewer high peaks in the GRFs). Overall, the $\Omega_m$-sensitivity of
the counts cannot be fully explained by the power spectrum.

\begin{figure*}
\begin{center}
\includegraphics[width=1.0\textwidth]{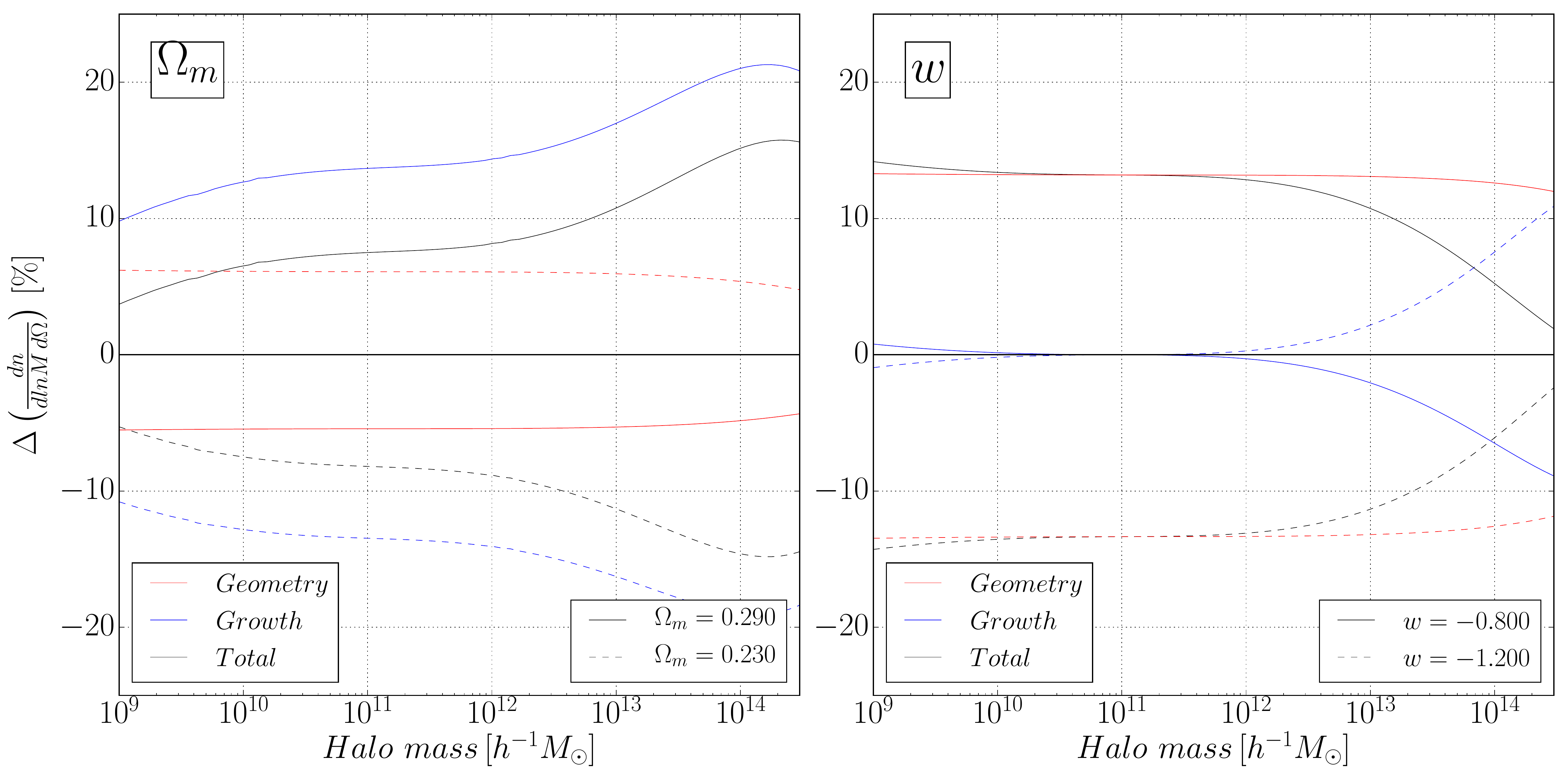}
\end{center}
\caption{(color online)\it Sensitivity of DM halo abundance to
  $\Omega_m$ (left panel) and $w$ (right panel). The percentage difference
  in the total number of halos per unit solid angle to $z=1$ between a
  model and the fiducial cosmology, as a function of the halo mass. The
  net effect (black) is decomposed into its geometry (red) and growth
  (blue) components.  } \label{halo}
\end{figure*}

To better understand the origin of the dependence of the peak counts'
sensitivity to peak height, we look at the 3D dark matter halo
counts. It is natural to compare these quantities, since high peaks
have long been known to be strongly correlated with individual
high-mass DM halos hosting galaxy clusters
\cite{White+2002,Hamana+2004,HennawiSpergel2005}.  The average number
of halos of a given mass to a fixed redshift per solid angle can be
expressed as an integral of the product of the volume element (geometry) and the 
halo mass function (growth).
\begin{eqnarray}
\frac{dn}{dlnM d\Omega}(M) = \int^{z_s}_0 dz \frac{dV}{dz d\Omega}(z) \frac{dn}{dlnM}(z,M)
\end{eqnarray}
We have computed the contribution from each effect as a function of
halo mass, and displayed the results in Fig.~\ref{halo}. The sensitivity
for halo masses above $\approx 10^{12} h^{-1} M_{\odot}$ tracks that
of high peaks, but this is not the case for low peaks / lower mass
halos. This is in agreement with previous studies that showed a link
between high peaks and single high-mass halos, while finding that
lower peaks are associated instead with constellations of 4-8 low-mass
halos at a range of redshifts \cite{Yang11}; a similar peak-halo
correlation has been seen in recent CFHTLens data~\cite{LiuHaiman2016}. High peaks then
seem to measure, like halos, measure a combination of growth and the volume element.

\begin{figure}
\begin{center}
\includegraphics[width=0.5\textwidth]{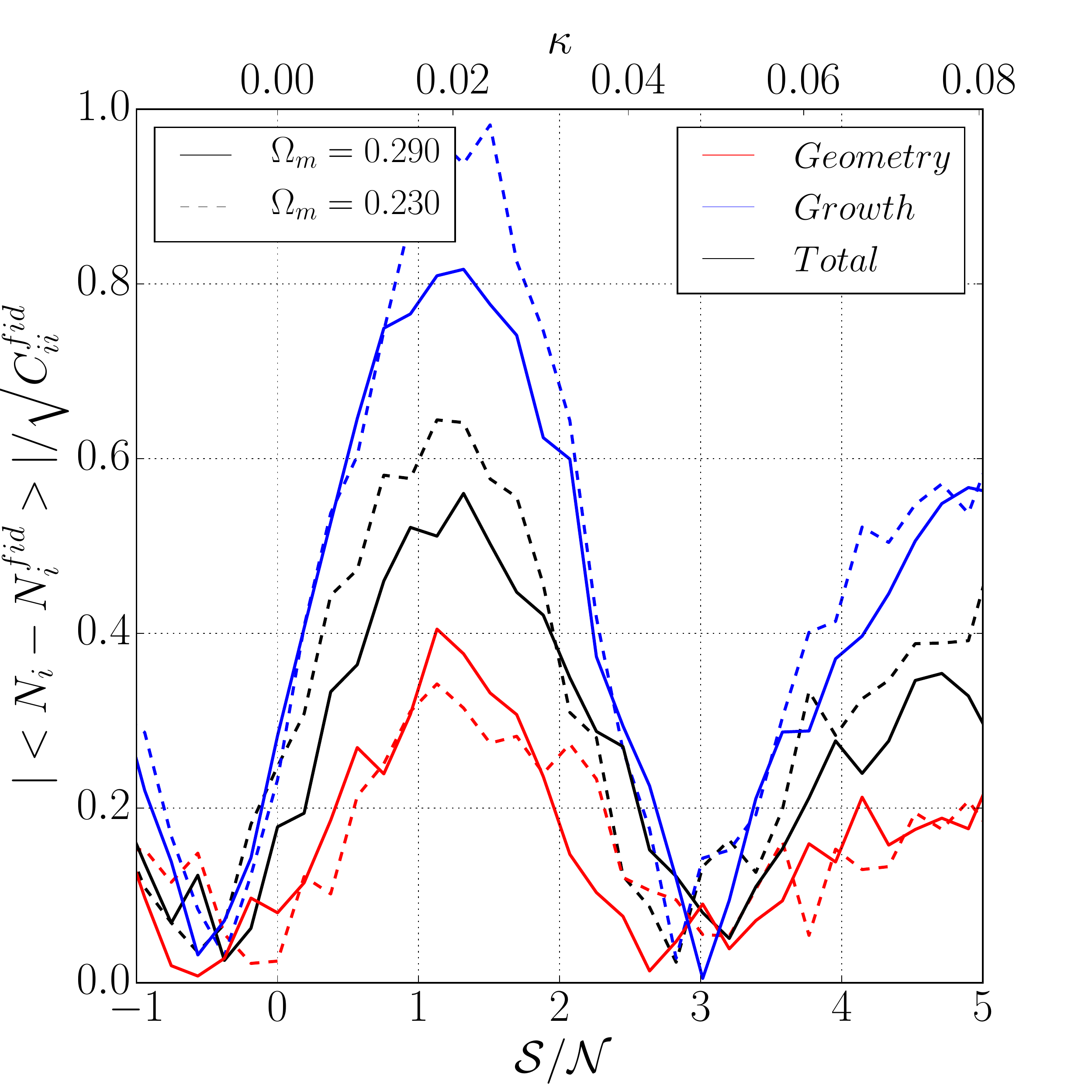}
\end{center}
\caption{(color online)\it 
Difference in number of peaks from the fiducial cosmology, 
normalized by the standard deviation in the fiducial model, 
for a 1000 deg2 survey.
} \label{peaksignificance}
\end{figure}

The sensitivity of the low peak counts does not track that of halo
counts; but these peaks are important for cosmology. When normalized
by the standard deviation for the fiducial model, the difference in peak counts from the
fiducial model has a maximum in the low significance region (see
Fig.~\ref{peaksignificance}). Low peaks have also been found to contribute
to cosmological parameter constraints more than high peaks, which is
in agreement with previous studies \cite{Yang11, Zorrilla16},
including an analysis of peak counts in the the CFHTLenS data
\cite{Liu15}.

The sensitivity of the Minkowski functionals, as well as its
decomposition into geometry and growth effects, qualitatively traces
that of lensing peaks, especially at high $\kappa$ levels.

\begin{figure}
\begin{center}
\includegraphics[width=0.5\textwidth]{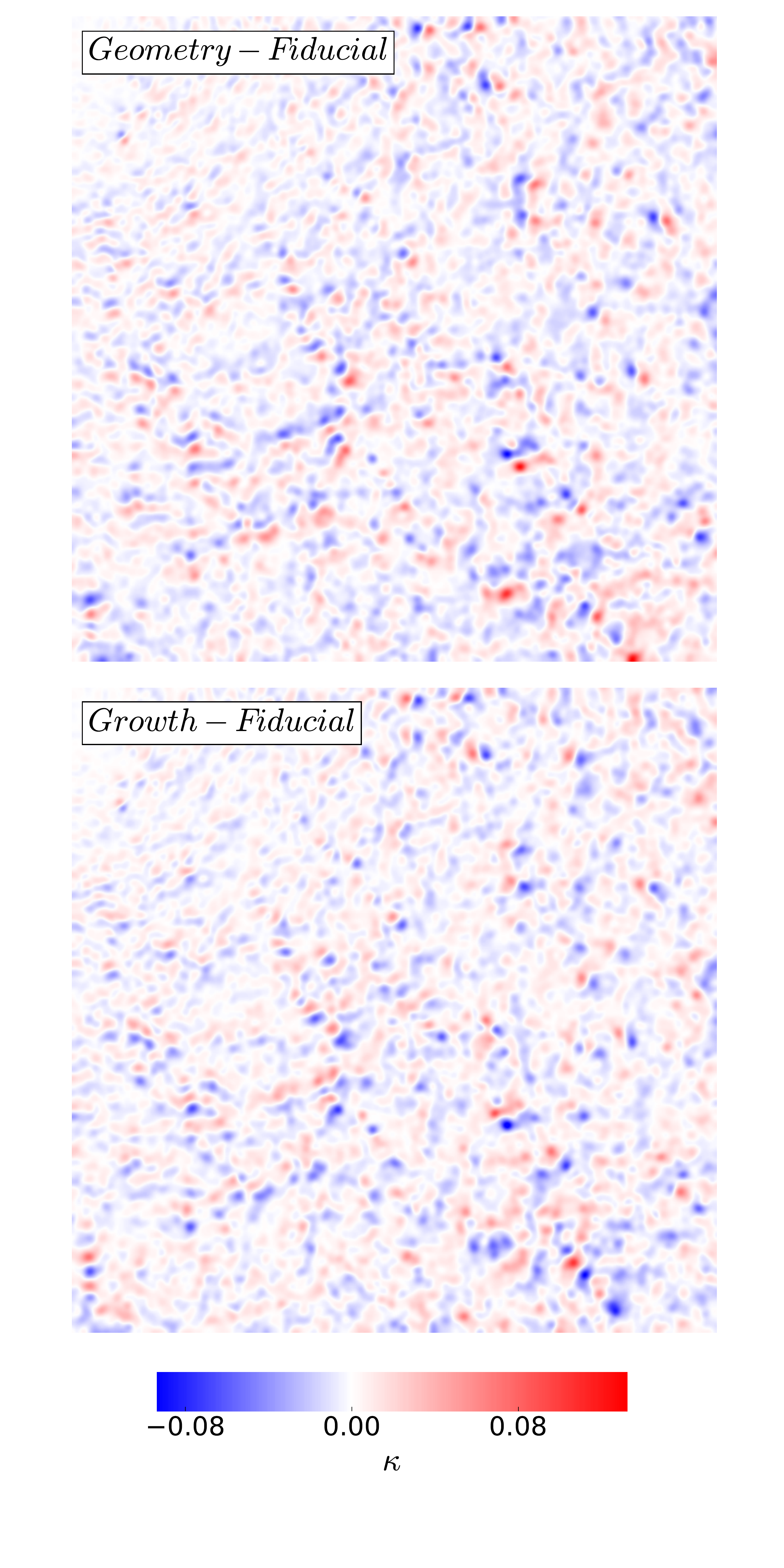}
\end{center}
\caption{(color online)\it Difference $\Delta\kappa$ between a single
  realization of the noisy $\kappa$ field generated including only
  geometry (top panel) or growth effects (bottom panel) for $w=-1.2$,
  and the corresponding realization in the fiducial model ($w=-1.0$).
} \label{maps}
\end{figure}

Finally, the fact that we observe a partial cancelation between geometry and
growth, especially when changing $w$, in all the statistics and
topological descriptors analyzed, suggests that this property is
present already at the map level. In order to investigate whether
this is the case, we have examined the difference-maps between each model
 and the fiducial, including either the geometry or growth effect alone. 
These maps are shown in Fig.~\ref{maps} for the model with $w=-1.2$. 
The modified angular positions of structures in the
maps built including each effect, due to different ray
deflections, prevent us from directly demonstrating a cancelation of
the lensing signal by adding these maps together. Nevertheless, the
geometric and growth-induced distortions in the two panels of
Fig.~\ref{maps} clearly show the same structures at roughly the same
locations, but with the sign of their $\Delta\kappa$ values reversed.
We conclude that the geometry {\it vs} growth cancelation indeed is a
property at the map level and we therefore expect it to affect any
observable, including those not analyzed here.

\section{Conclusions}\label{conclusions}

We have validated the use of N-body simulations and ray-tracing to
separately study the effect of geometry and growth on weak lensing
observables. This allows us to extend past analyses to non-Gaussian
statistics and topological features that do not admit a simple
analytic treatment.

Our analysis confirms that the sensitivity of non-Gaussian observables
to cosmology shares some characteristics with that of the power
spectrum. They suffer a partial cancellation between geometry and growth
on top of the loss of sensitivity due to integrating (projection) effects. This 
cancellation is more severe for $w$, reducing even further the sensitivity of
WL to that parameter compared to $\Omega_m$.

Galaxy shape noise dominates the power spectrum at high multipoles,
reinforcing the case to use alternative observables to analyze weak
lensing data on small scales. The bispectrum has higher statistical noise,
but shape noise does not suppress its average sensitivity at high multipoles
as it does for the power spectrum. The lensing peaks' sensitivity is
highly dependent on the peak height, with high peaks tracking the
behavior of dark matter halo counts, but low peaks having an important
influence on parameter constraints. The sensitivity of Minkowski
functionals is similar to that of peak counts, which is not surprising at high $\kappa$ levels. 
The similarities between statistics, such as the cancelation of geometry and growth
effects, arises from the fact that this property is present at map
level.

The partial cancelation, together with projection effects, yields weak constrains for 
$w$, and underscores the need to combine information from
different redshifts to tighten constrains on DE. Marginalized errors
on $\Omega_m$ and $w$ are similar to those calculated from growth-only
effects. This suggests that combining WL data with probes that
strongly constrain the expansion history through geometry, such as BAO
measurements, may be especially beneficial to tighten constraints.

\section*{Acknowledgments}

We acknowledge stimulating conversations with Lloyd Knox, which
motivated this work. We thank the anonymous referee for useful comments 
that improved this paper. The ray--tracing simulations and parameter
estimation calculations were performed at the NSF XSEDE facility,
supported by grant number ACI-1053575.  This work was supported in
part by NSF Grant No. AST-1210877, the Research Opportunities and
Approaches to Data Science (ROADS) program at the Institute for Data
Sciences and Engineering at Columbia University.  ZH gratefully
acknowledges support by a Simons Fellowship in Theoretical Physics.

\clearpage

\bibliography{ref}

\end{document}